\documentclass[epj,nopacs]{svjour}
\usepackage{graphics}
\def\gsim{\raise0.3ex\hbox{$>$\kern-0.75em\raise-1.1ex\hbox{$\sim$}}}
\def\lsim{\raise0.3ex\hbox{$<$\kern-0.75em\raise-1.1ex\hbox{$\sim$}}}
\begin{document}
\title{Direct photons measured by the PHENIX experiment at RHIC}
\author{Stefan Bathe\thanks{\emph{Present address:} Brookhaven National Lab,Bldg. 510 C, Upton, NY 11973, USA} 
%
for the PHENIX Collaboration
}                     
\offprints{}          
\institute{University of California at Riverside}
\date{Received: date / Revised version: date}
%
\abstract{ Results from the PHENIX experiment at RHIC on direct photon
production in $p+p$, $d+$Au, and Au+Au collisions at
$\sqrt{s_{\scriptscriptstyle NN}}$ = 200 GeV are presented.  In $p+p$
collisions, direct photon production at high $p_T$ behaves as expected
from perturbative QCD calculations.  The $p+p$ measurement serves as a
baseline for direct photon production in Au+Au collisions. In $d+$Au
collisions, no effects of cold nuclear matter are found within the
large uncertainty of the measurement.  In Au+Au collisions, the
production of high $p_T$ direct photons scales as expected for
particle production in hard scatterings.  This supports jet quenching
models, which attribute the suppression of high $p_T$ hadrons to the
energy loss of fast partons in the medium produced in the collision.
Low $p_T$ direct photons, measured via $e^+e^-$ pairs with small
invariant mass, are possibly related to the production of thermal
direct photons.
\PACS{
      {PACS-key}{discribing text of that key}   \and
      {PACS-key}{discribing text of that key}
     } 
} 
\maketitle
\section{Introduction}
\label{intro}
Depending on their transverse momentum, direct photons convey
information about different aspects of ultra-relativistic
nucleus-nucleus ($A+A$) collisions.  High transverse momentum ($p_T$)
direct photons are produced in early, hard parton-parton scatterings
by processes like quark-gluon Compton scattering (q+g
$\longrightarrow$ q + $\gamma$).  Unlike scattered quarks or gluons,
direct photons do not interact strongly with the medium subsequently
produced in the collision.  High $p_T$ direct photons ($p_T \gsim 6$
GeV/$c$) therefore provide a baseline for measuring medium
modifications of high $p_T$ hadron production.

A significant fraction of low $p_T$ direct photons ($1 \leq p_T \leq 3$
GeV/$c$ in central Au+Au collisions at $\sqrt{s_{\scriptscriptstyle
NN}}$ = 200 GeV) is expected to come from a thermalized medium of
deconfined quarks and gluons, the Quark-Gluon Plasma (QGP), possibly
created in $A+A$ collisions \cite{Turbide:2003si}.  These thermal
photons carry information about the initial temperature of the QGP.
Thermal photons are produced in the QGP as well as in the hadronic gas
over the entire life time of the collision.  The initial temperature
can be extracted comparing measurements to models which convolute the
production rate with the space-time evolution of the collision.

At low and intermediate $p_T$ ($p_T \leq 6$ GeV/$c$ in central Au+Au
collisions at $\sqrt{s_{\scriptscriptstyle NN}}$ = 200 GeV) another
significant source of direct photons might be the interaction of fast
partons from jets with thermal partons in the QGP
\cite{Gale:2005zd} like ${\mathrm q}_{\mathrm hard} + {\mathrm
g}_{\mathrm QGP} \longrightarrow {\mathrm q} + \gamma$.

Possible modifications of direct photon production from cold nuclear
matter effects can be measured in $d+$Au collisions where no medium is
created.  Direct photon measurements in $p+p$ collisions are a superb
test of QCD.  They constrain the gluon distribution function since the
gluon is a direct participant of the partonic scattering.
Furthermore, they provide a baseline for understanding direct photon
production in $A+A$ collisions.

\section{High \boldmath $p_T$ \unboldmath direct photons}
\label{sec_high}

  \subsection{Measurement}
  \label{subsec_high_meas}

  Experimentally, the measurement of direct photons is challenging due
  to a large background from hadron decays like $\pi^0 \longrightarrow
  \gamma \gamma$ and $\eta \longrightarrow \gamma \gamma$.  PHENIX
  measures all those photon sources.  Photons are measured with the
  electromagnetic calorimeter (EMCal).  Neutral pions and $\eta$
  mesons are measured through their two-photon decay branch
  \cite{Adler:2003qi,Adler:2006hu}.  The EMCal
  \cite{Aphecetche:2003zr} consists of six sectors of a lead
  scintillator and two sectors of a lead glass calorimeter centered at
  midrapidity ($\eta < 0.35$) and covering 1/4 of the azimuthal angle.
  
  The direct photon spectrum is obtained by subtracting the decay
  photon spectrum from the inclusive photon spectrum.  This is gained
  from photon-like showers in the EMCal corrected for contaminations
  from charged hadrons and neutrons.  The decay photon spectrum is
  calculated from the measured $\pi^0$ and $\eta$ spectrum, taking
  into account minor contributions from other hadrons that decay into
  photons.

  If the detector occupancy is low as in $p+p$ or $d+$Au collisions,
  the signal-to-background ratio can be improved by event by event
  tagging of photons that have a matching partner as decay photons.

  \subsection{Results}
  \label{subsec_high_res}

    \subsubsection{$p+p$ collisions}
    \label{subsubsec_pp}

    \begin{figure}
      \resizebox{0.45\textwidth}{!}{
	\includegraphics{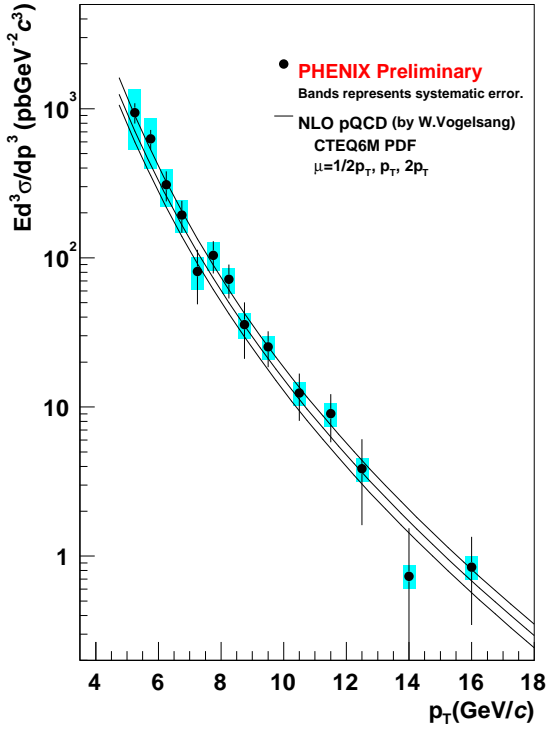}}
      \caption{Preliminary direct photon cross section as a function
	of $p_T$ for $p+p$ collisions at $\sqrt{s}$ = 200 GeV
	\cite{Okada:2005in}.  The solid curves are pQCD predictions
	\cite{Gordon:1993qc} for three different scales, which
	represent the uncertainty on the calculation.}
      \label{fig:fig1_pp}
    \end{figure}

    The preliminary direct photon cross section for $p+p$ collisions
    at $\sqrt{s}$ = 200 GeV is shown in Fig.  \ref{fig:fig1_pp}
    \cite{Okada:2005in}.  It is based on an integrated luminosity of
    266 nb$^{-1}$ collected in RHIC Run-3.  For the whole $p_T$ range
    of 5 to 16 GeV/$c$ the observed cross section is consistent with a
    next-to-leading-order perturbative-QCD (NLO pQCD) calculation
    \cite{Gordon:1993qc}.  This measurement establishes a reference
    for direct photon production in $A+A$. In RHIC Run-5, PHENIX took
    a $p+p$ data set with ten times the Run-3 statistics.

    \subsubsection{Au+Au collisions}
    \label{subsubsec_AuAu}

    The high $p_T$ direct photon yield in $A+A$ collisions relative to
    $p+p$ is expected to scale with the parton luminosity in the
    overlap region of the two nuclei.  The parton luminosity is
    quantified by the nuclear overlap function, $T_{AA}$.  In the
    absence of nuclear effects the nuclear modification factor
    \begin{equation}
    R_{AA}(p_T) = \frac{{\mathrm d}N/{\mathrm d}p_T|_{\mathrm A+A}}
        {\langle T_{AA} \rangle_{f} \times {\mathrm d}\sigma/{\mathrm d}p_T|
        _{p+p}}
    \end{equation}
    is unity for particle production from hard scattering.

\begin{figure}
      \resizebox{0.45\textwidth}{!}{
	\includegraphics{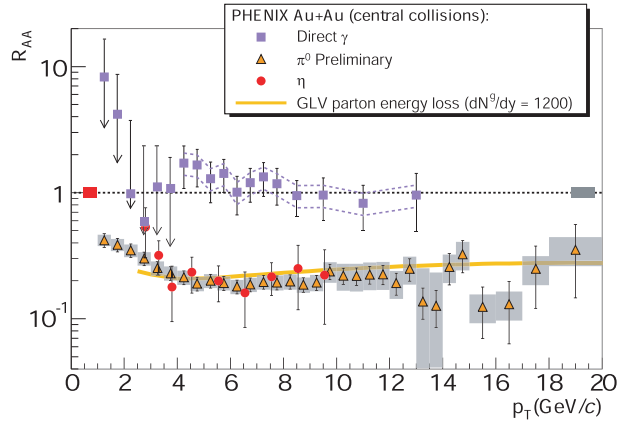}}
      \caption{Nuclear modification factor for direct photon, $\pi^0$,
               and $\eta$ production in central Au+Au collisions at
               $\sqrt{s_{\scriptscriptstyle NN}}$ = 200 GeV.  For
               direct photons, a pQCD calculation is used as the
               reference.}
      \label{fig:fig3_RAA_comp}
    \end{figure}

    As shown by PHENIX in RHIC Run-2 \cite{Adler:2005ig}, at high
    $p_T$ direct photon production is consistent with the $T_{AA}$-scaled pQCD
    expectation also in Au+Au collisions.  Figure
    \ref{fig:fig3_RAA_comp} compares the nuclear modification factor
    for direct photon, $\pi^0$, and $\eta$ production in central Au+Au
    collisions at $\sqrt{s_{\scriptscriptstyle NN}}$ = 200 GeV.  For
    direct photons, the pQCD calculation is used as a reference.  The
    suppression for $\pi^0$ and $\eta$, supported by the
    non-suppression of photons, can be described with energy loss of
    partons in the produced medium.

    While high $p_T$ direct photon production is consistent with the
    pQCD expectation, one should keep in mind that there are a number
    of medium effects that might alter direct photon production.  On
    the one hand, a significant fraction of direct photons is expected
    to stem from the fragmentation of hard-scattered partons into
    jets.  As those partons lose energy in the medium, also the
    fragmentation photons should be suppressed.  On the other hand, as
    partons lose energy through gluon bremsstrahlung, they should also
    generate photon bremsstrahlung, enhancing the direct photon yield.
    It would be a coincidence if these counterbalancing effects
    exactly canceled.
    
    Experimentally, contributions from these effects can be quantified
    by studying the azimuthal distribution of direct photon production
    with respect to the reaction plane.  If fragmentation photons are
    suppressed, less direct photons are expected to be emitted outside
    of the reaction plane.  If bremsstrahlung is enhanced, more direct
    photons will be found outside of the reaction plane.

    \subsubsection{$d+$ Au collisions}
    \label{subsubsec_dAu}

\begin{figure}
      \resizebox{0.45\textwidth}{!}{
	\includegraphics{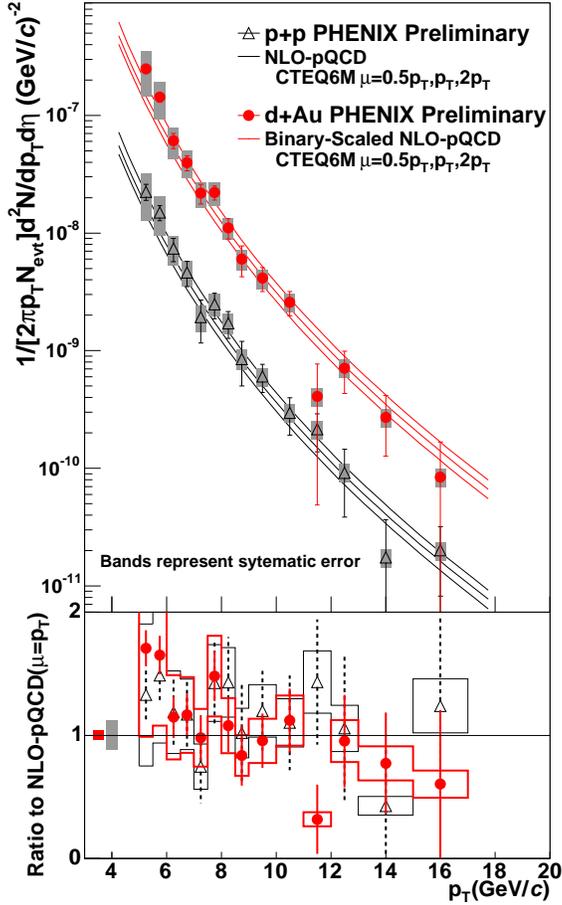}}
      \caption{Preliminary direct photon invariant yield as a function
	of $p_T$ for $p+p$ \cite{Okada:2005in} and minimum-bias $d$+Au
	collisions at $\sqrt{s_{\scriptscriptstyle NN}}$ = 200 GeV.
	The solid curves are pQCD predictions \cite{Gordon:1993qc} for
	three different scales, which represent the uncertainty on the
	calculation.}
      \label{fig:fig2_ppdAu}
    \end{figure}

    The preliminary direct photon yield for minimum-bias $d$+Au
    collisions at $\sqrt{s_{\scriptscriptstyle NN}}$ = 200 GeV is
    shown in the upper panel of Fig.  \ref{fig:fig2_ppdAu} along with
    the yield in $p+p$ collisions.  The $d+$Au analysis is based on
    $\approx$ 3 billion events sampled in RHIC Run-3.  The lower panel
    shows the ratio to the $T_{AA}$-scaled NLO pQCD calculation.  The
    ratio is consistent with unity over the entire $p_T$ range,
    showing no indication of cold nuclear matter effects.  However,
    the uncertainty of the measurement is large.  A high statistics
    $d+$Au run is planned.

\section{Low \boldmath $p_T$ \unboldmath direct photons}
\label{sec_low}

  \subsection{Measurement}
  \label{subsec_low_meas}

  The EMCal measurement at low $p_T$ suffers from a large background
  from decay photons.  Also, the relative energy resolution of the
  EMCal becomes worse at low energy.  Therefore, a new method
  \cite{Cobb:1978gj,Albajar:1988iq} to measure direct photons at low
  $p_T$ has been employed.  It has been carried out in heavy ion
  experiments for the first time. The method is based on the
  measurement of pairs of $e^+e^-$, which are identified with the
  PHENIX Ring Imaging Cherenkov Detector (RICH).  The basic idea is
  that any source of real photons also produces virtual photons that
  decay into $e^+e^-$ with small invariant mass.  An example is the
  $\pi^0$ Dalitz decay ($\pi^0 \longrightarrow \gamma e^+e^-$).  This
  internal conversion method is based on two assumptions: first, the
  ratio of direct-to-all photons is the same for real and virtual
  photons at small invariant mass close to zero ($m_{\gamma} < 30
  $MeV): $\gamma_{\mathrm direct}^\star/\gamma_{\mathrm incl.}^\star =
  \gamma_{\mathrm direct}/\gamma_{\mathrm incl.}$.  Second, the mass
  distribution can be described by the Kroll-Wada formula
  \cite{Kroll:1955zu}, which has been established to describe the
  Dalitz decay
  \footnote{Since the virtual photons decay in the medium, this
  relation might be slightly modified.  This would not affect the
  significance of the observed excess of direct photons, only its
  translation into an absolute yield of real direct photons.}:

  \begin{eqnarray}
    \lefteqn{\frac{1}{N_\gamma} \frac{dN_{ee}}{dm_{ee}}=}  \\ 
      & & \frac{2\alpha}{3\pi}
    \sqrt{1 - \frac{4m_e^2}{m_{ee}^2}} (1 + \frac{2m_e^2}{m_{ee}^2}) 
    \frac{1}{m_{ee}} \mid F(m_{ee}^2) \mid^2 (1 - \frac{m_{ee}^2}{M^2})^3 \; ,
     \nonumber
    \label{eq:Wa}
  \end{eqnarray}

  \begin{figure}
    \resizebox{0.45\textwidth}{!}{
	\includegraphics{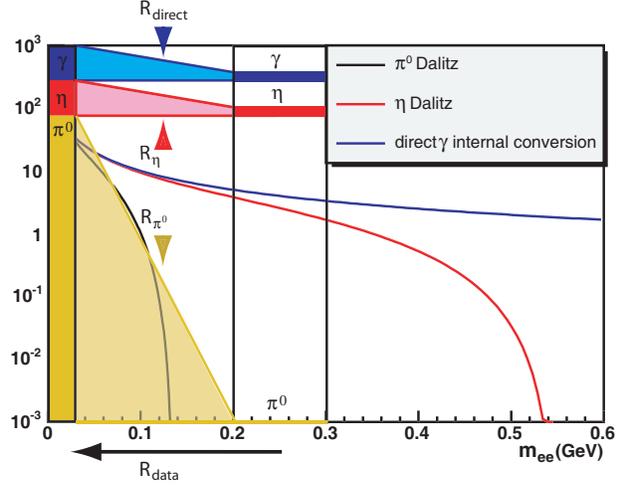}}
    \caption{Invariant-mass distribution of virtual photons from the
    $\pi^0$ and $\eta$ Dalitz decay as well as from direct photons
    according to Eq. \ref{eq:Wa}.  It is illustrated how the various
    contributions decrease to a fraction $R$ when going to higher
    invariant mass with the $\pi^0$ contribution exhausting.}
    \label{fig:fig4_sketch}
  \end{figure}
  
  The mass distribution according to Eq. \ref{eq:Wa} is depicted in
  Fig. \ref{fig:fig4_sketch}.  For $e^+e^-$ pairs from $\pi^0$ and
  $\eta$ Dalitz decays the yield is suppressed towards higher $m_{ee}$
  due to the mass, $M$, of the parent meson in the phase space factor
  $(1-m^2_{ee}/M^2)^3$ whereas no suppression takes place for $e^+e^-$
  pairs from virtual direct photons as long as $m_{ee} \ll p_T^{ee}$.
  For the small invariant masses considered here the form factor 
  $|F(m^2_{ee})|$ is assumed to be unity in all cases.

  The key advantage of this method is the greatly improved
  signal-to-background ratio, which is achieved by eliminating the
  contribution from $\pi^0$ Dalitz decays when the invariant mass is
  increased.  In addition, the electron measurement through charged
  particle tracking provides a better energy resolution at low $p_T$
  than the EMCal photon measurement.  This excellent energy resolution
  at low $p_T$ combined with little material upstream of the detector,
  where photons could convert generating background $e^+e^-$ pairs,
  makes this measurement feasible in PHENIX.

  The experimentally observed quantity is the ratio of $e^+e^-$ pairs
  in an invariant mass bin where the $\pi^0$ Dalitz decay is largely
  suppressed to the yield at an invariant mass close to zero:
  $R_\mathrm{data} = N^\mathrm{90-300 MeV}_{ee}/N^\mathrm{<30
  MeV}_{ee}$.  If there is no direct photon signal, $R_\mathrm{data} =
  R^\mathrm{calc}_\mathrm{hadron}$, {\it i.e.} $R_\mathrm{data}$ can
  be calculated\footnote{Besides $\pi^0$ and $\eta$, decay photons
  from a cocktail of hadrons are considered here.} from
  Eq. \ref{eq:Wa} based on the known ratio $\eta/\pi^0 = 0.45 \pm
  0.1$.  An excess $R_\mathrm{data} > R^\mathrm{calc}_\mathrm{hadron}$
  translates into the fraction of virtual direct photons at close to
  zero invariant mass according to:
  \begin{equation}
    \frac{\gamma_{\mathrm direct}^\star}{\gamma_{\mathrm incl.}^\star} =
     \frac{R_\mathrm{data}-R^\mathrm{calc}_\mathrm{hadron}}
      {R^\mathrm{calc}_{\mathrm{direct} \, \gamma}-R^\mathrm{calc}_\mathrm{hadron}}
  \end{equation}
  This can be derived starting from 
  \begin{equation}
    R_\mathrm{data} = N^\mathrm{90-300 MeV}_{ee}/N^\mathrm{<30 MeV}_{ee}
  \end{equation}
  with 
  \begin{equation}
    N_{ee} = N_\mathrm{hadron} + N_{\mathrm{direct} \, \gamma}:
  \end{equation}
  \begin{eqnarray}
    R_\mathrm{data} & = & \frac{R^\mathrm{calc}_\mathrm{hadron} N^\mathrm{<30 MeV}_\mathrm{hadron} + R^\mathrm{calc}_{\mathrm{direct} \, \gamma} N^\mathrm{<30 MeV}_{\mathrm{direct} \, \gamma}}{N^\mathrm{<30 MeV}_\mathrm{hadron} + N^\mathrm{<30 MeV}_{\mathrm{direct} \, \gamma}} \\
                    & = & R^\mathrm{calc}_\mathrm{hadron} + (R^\mathrm{calc}_{\mathrm{direct} \, \gamma} - R^\mathrm{calc}_\mathrm{hadron}) \frac{N^\mathrm{<30 MeV}_{\mathrm{direct} \, \gamma}}{N^\mathrm{<30 MeV}_{ee}}
  \end{eqnarray}
  The direct photon spectrum is then obtained by multiplying $\gamma_{\mathrm direct}^\star/\gamma_{\mathrm incl.}^\star$
  by the inclusive photon spectrum measured with the EMCal.

  The full 2004 data set of about 900 M minimum bias events is
  analyzed.  Events and centrality are selected as described in
  \cite{Adcox:2003nr}.  Electrons in the central arms are identified
  by matching charged particle tracks to clusters in the EMCal and to
  rings in the RICH detector.  To obtain a clean invariant-mass
  distribution of $e^+e^-$ pairs, pairs originating from photon
  conversions in the beam pipe or detector material are rejected based
  on their orientation with respect to the magnetic field.  The
  combinatorial background is removed by an event-mixing technique.
  The uncertainty of the $\eta$-to-$\pi^0$ ratio of about 20 \%
  \cite{Adler:2004ta} is the main source of uncertainty, translating
  into an uncertainty of 20 \% of the measured direct photon yield.
  Other sources are the EMCal-measured inclusive photon yield (10 \%)
  and the $e^+e^-$-pair acceptance (5~\%).  The total systematic
  uncertainty is 25 \%.

  \subsection{Results}
  \label{subsec_low_res}

  \begin{figure*}
    \begin{center}
	\resizebox{0.75\textwidth}{!}{
	  \includegraphics{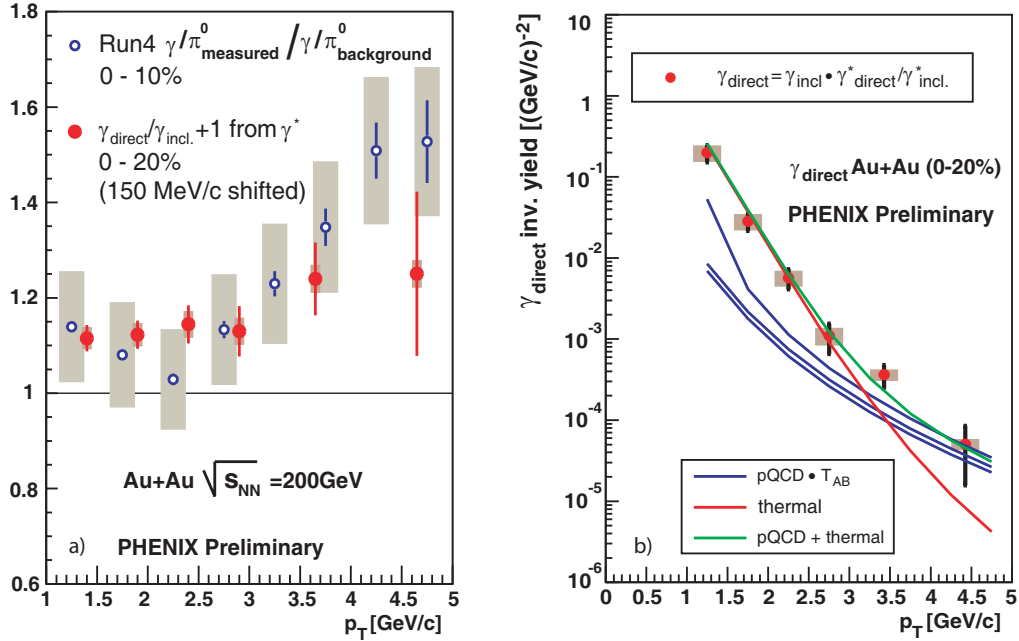}}
    \end{center}
    \caption{a) Direct photon excess for the conventional and the
    internal-conversion measurement in central Au+Au collisions
    \cite{Bathe:2005nz}.  b) Direct photon spectrum from the latter
    \cite{Bathe:2005nz} compared to pQCD \cite{Gordon:1993qc},
    thermal-photon \cite{d'Enterria:2005vz}, and the sum of both
    calculations.}
    \label{fig:fig5_ratio_Aspe}
  \end{figure*}

  Figure \ref{fig:fig5_ratio_Aspe} a) shows the direct photon signal
  for central Au+Au collisions at $\sqrt{s_{\scriptscriptstyle NN}}$ =
  200 GeV in terms of the double ratio 
  $(\gamma/\pi^0|_\mathrm{meas.})/(\gamma/\pi^0|_\mathrm{bckgrd.})$.  
  This ratio indicates a
  direct photon excess above the decay photon contribution as an
  enhancement above 1. In this figure, two measurements are compared:
  the result from the internal conversion method as described above
  \cite{Bathe:2005nz} and a preliminary analysis of the conventional
  EMCal measurement based on a subset of the RHIC Run-4 data set
  \cite{Bathe:2005nz}.
  \footnote{The preliminary result from the measurement of virtual
  photons is presented in the same figure in terms of
  $\gamma_{\mathrm direct}/\gamma_{\mathrm incl.} + 1$, which also
  indicates a direct photon excess as an enhancement above 1. Note
  that the two quantities are not exactly equivalent.}  While the
  internal conversion measurement results in a significant direct
  photon signal of about 10 \% above the decay photon background for
  $1 < p_T < $ 5 GeV/$c$, the EMCal measurement does not yield a
  significant direct photon signal below $p_T = $ 3 GeV/$c$, but
  agrees with the internal conversion measurement within the
  uncertainties over the entire range.

  Figure \ref{fig:fig5_ratio_Aspe} b) shows the direct photon
  invariant yield from the internal conversion
  measurement \cite{Bathe:2005nz} and compares it to theoretical
  calculations.  With large significance, a direct photon spectrum is
  obtained for $1 < p_T < $ 5 GeV/$c$.  The spectrum lies
  significantly above a $T_{AA}$-scaled NLO pQCD calculation
  \cite{Gordon:1993qc} for $p_T$ \lsim 3 GeV/$c$.  The pQCD
  calculation indicates the contribution from hard scatterings.  The
  lines show the scale uncertainty of the calculation.  However,
  it is not clear how meaningful the comparison is down to this low
  $p_T$ where the pQCD calculation reaches its limit of applicability.
  It is planned to replace the pQCD calculation by a reference
  measurement of direct photon production in $p+p$ collisions with the
  same method as in Au+Au.

  The excess of the measured direct photon spectrum above the pQCD
  calculation can be described by models that allow for a significant contribution of thermal photons.
  In order to describe thermal photon production in $A+A$
  collisions, the entire space-time evolution has to be accounted for
  including the hadronic phase.  This is done in hydrodynamical
  models, which assume local thermal equilibrium.  An important free
  parameter in such models is the initial temperature of the fireball.
  Figure \ref{fig:fig5_ratio_Aspe} b) compares the measured direct
  photon spectrum to a 2+1 hydrodynamical model
  \cite{d'Enterria:2005vz} for thermal-photon emission with an average
  initial temperature of $T_0^{\mathrm ave}=$ 360 MeV ($T_0^{\mathrm
  max}=$ 570 MeV) and a formation time of $\tau_0=$ 0.15 fm/$c$.  The
  model underpredicts the data for $p_T$ \gsim 3 GeV/$c$.  The data
  can be described when both sources, thermal and pQCD, are combined.
  Various other calculations come to similar descriptions with initial
  temperatures in the range $370 < T_i < 570$ MeV
  \cite{Gale:2005zd,Alam:2005za,Chaudhuri:2005dp}.  The temperatures
  are significantly above the critical temperature for the QGP phase
  transition of $T_c$ = 170 MeV.  However, the obtained temperature is
  only meaningful if the excess above pQCD is of thermal origin.
  Besides hard photons, also jet-plasma interactions may play a
  significant role \cite{Gale:2005zd}.

\section{Conclusions}
\label{sec_conc}

The direct photon cross section in $p+p$ collisions at $\sqrt{s}$ =
200 GeV is found to be consistent with a NLO pQCD calculation over
the entire range of the measurement from 5 GeV $ \le p_T \le 16$
GeV/$c$.  In $d+$Au collisions, no indications for cold nuclear
matter effects are found within the large uncertainties of the
measurement.  A high statistics $d+$Au run is planned.  
High $p_T$ direct photon production in Au+Au collisions is
found to scale with the nuclear overlap function $T_{AA}$,
establishing medium effects as the cause for the hadron suppression
observed in central Au+Au collisions.

At low $p_T$ ($1 \le p_T \le 5$ GeV/$c$) a significant direct photon
excess above the decay photon background is observed employing a new
method in heavy ion collisions, which measures low invariant mass
$e^+e^-$ pairs from direct photon internal conversions.  The signal
appears to be above pQCD calculations for $p_T \le 3$ GeV/$c$, whose
applicability is questionable, however, at this low $p_T$.  If a
thermal photon source modeled by hydrodynamical calculations is added
to the pQCD calculation, the data can be described over the entire
$p_T$ range.  Other direct photon sources like jet-plasma interactions
might play a significant role as well.

\bibliographystyle{unsrt.bst}
\bibliography{HQ06ProcBathe}

\end{document}